\documentclass[aps,pre,groupedaddress,superscriptaddress,reprint,showpacs,twocolumn,amsmath,amssymb]{revtex4}

\usepackage[]{graphicx}
\usepackage{amsmath}
\usepackage{enumerate}
\usepackage{subeqnarray}
\usepackage{cases}

\begin{document}

\title{Statistical mechanics for non-reciprocal forces}

\author{A. V. Ivlev}
\email[e-mail:~]{ivlev@mpe.mpg.de} \affiliation{Max-Planck-Institut f\"ur Extraterrestrische Physik, Giessenbachstr., 85741
Garching, Germany}
\author{J. Bartnick}
\author{M. Heinen}
\author{H. L\"owen}
\affiliation{Institut f\"ur Theoretische Physik II, Weiche Materie, Heinrich-Heine-Universit\"at, 40225 D\"usseldorf,
Germany}

\pacs{05.20.-y, 52.27.Lw, 65.20.De, 05.20.Dd}

\begin{abstract}
A basic statistical mechanics analysis of many-body systems with non-reciprocal pair interactions is presented. Different
non-reciprocity classes in two- and three-dimensional binary systems (relevant to real experimental situations) are
investigated, where the action-reaction symmetry is broken for the interaction between different species. The asymmetry is
characterized by a non-reciprocity parameter $\Delta$, which is the ratio of the non-reciprocal to reciprocal pair forces.
It is shown that for the ``constant'' non-reciprocity (when $\Delta$ is independent of the interparticle distance $r$) one
can construct a pseudo-Hamiltonian and such systems, being intrinsically non-equilibrium, can nevertheless be described in
terms of equilibrium statistical mechanics and exhibit detailed balance with distinct temperatures for the different
species. For a general case (when $\Delta$ is a function of $r$) the temperatures grow with time, approaching a universal
power-law scaling, while their ratio is determined by an {\it effective} constant non-reciprocity which is uniquely defined
for a given interaction.
\end{abstract}

\maketitle

One of the fundamental postulates of classical mechanics is Newton's third law {\it actio}={\it reactio}, which states that
the pair interactions between particles are reciprocal. Newton's third law holds both for the fundamental microscopic
forces, but also for {\it equilibrium} effective forces on classical particles, obtained by integrating out microscopic
degrees of freedom \cite{Israelachvili,Dijkstra00,Bolhuis01,Praprotnik08,Mognetti09}. However, the action-reaction symmetry
for particles can be broken when their interaction is mediated by some {\it non-equilibrium environment}: This occurs, for
instance, when the environment moves with respect to the particles, or when a system of particles includes different species
and their interaction with the environment is out of equilibrium (of course, Newton's third law holds for the complete
``particles plus environment'' system). Examples of non-reciprocal interactions on the mesoscopic length-scale include
forces induced by non-equilibrium fluctuations \cite{Hayashi06,Buenzli08}, optical \cite{Dholakia10} and diffusiophoretic
\cite{Sabass10,Soto14} forces, effective interactions between colloidal particles under solvent or depletant flow
\cite{Dzubiella03,Khair07,Mejia11,Sriram12}, shadow \cite{Tsytovich97,Khrapak01,Chaudhuri11} and wake-mediated
\cite{Melzer99,Morfill09,Couedel10} interactions between microparticles in a flowing plasma, etc. A very different case of
non-reciprocal interactions are ``social forces'' \cite{Helbing95,Helbing00} governing, e.g., pedestrian dynamics.

Non-reciprocal forces are in principle non-Hamiltonian, so the standard Boltzmann description of classical equilibrium
statistical mechanics breaks down. Hence, it is a priori unclear whether concepts like temperature and thermodynamic phases
can be used to describe them. To the best of our knowledge, the classical statistical mechanics of systems with
non-reciprocal interactions -- despite their fundamental importance -- has been unexplored so far.

In this Letter we present the statistical foundations of systems with non-reciprocal interparticle interactions. We consider
a binary system of particles, where the action-reaction symmetry is broken for the pair interaction between different
species. The asymmetry is characterized by the non-reciprocity parameter $\Delta$, which is the ratio of the non-reciprocal
to reciprocal forces. We show that for the ``constant'' non-reciprocity, when $\Delta$ is independent of the interparticle
distance $r$, one can construct a (pseudo) Hamiltonian with renormalized masses and interactions. Hence, being intrinsically
non-equilibrium, such systems can nevertheless be described in terms of equilibrium statistical mechanics and exhibit
detailed balance with distinct temperatures for different species (the temperature ratio is determined by $\Delta$). For a
general case, when $\Delta$ is a function for $r$, the system is no longer conservative -- it follows a universal asymptotic
behavior with the temperatures growing with time as $\propto t^{2/3}$. The temperature ratio in this case is determined by
an {\it effective} constant non-reciprocity which is uniquely defined for a given interaction. In the presence of frictional
dissipation the temperatures reach a steady state, while their ratio remains practically unchanged.

Let us consider a binary mixture of particles of the sort $A$ and $B$. The spatial dependence of the pair interaction is
described by the function $\varphi(r)$. The interaction is reciprocal for the $AA$ and $BB$ pairs, while between the species
$A$ and $B$ the action-reaction symmetry is broken. The measure of the asymmetry is the {\it non-reciprocity parameter}
$\Delta(\geq0)$ which we first assume to be independent of the interparticle distance (``constant''). We present the force
${\bf F}_{ij}$ exerted by the particle $i$ on the particle $j$ as follows:
\begin{equation}\label{force_const}
{\bf F}_{ij} = -\frac{\partial\varphi(r_{ij})}{\partial{\bf r}_{j}}\times\left\{
\begin{array}{cl}
1-\Delta & \text{ for $ij \in AB$};\\
1+\Delta & \text{ for $ij \in BA$};\\
1 & \text{ for $ij \in AA$ or $BB$},
\end{array}
\right.
\end{equation}
where $r_{ij}=|{\bf r}_i-{\bf r}_j|$ and each particle can be of the sort $A$ or $B$; note that $\varphi(r)$ may be
different for different pairs \cite{note0}. By writing the Newtonian equations of motion of individual particles interacting
via the force (\ref{force_const}), we notice that the interaction symmetry is restored if the particle masses and
interactions are renormalized as follows:
\begin{eqnarray}\label{mass_ren}
\tilde m_{i} &=& m_i\times\left\{
\begin{array}{rl}
(1+\Delta)^{-1} & \text{ for $i\in A$};\\
(1-\Delta)^{-1} & \text{ for $i\in B$},
\end{array}
\right.\\[.3cm]\label{phi_ren}
\tilde \varphi(r_{ij}) &=& \varphi(r_{ij})\times\left\{
\begin{array}{cl}
(1+\Delta)^{-1} & \text{ for $ij\in AA$};\\
(1-\Delta)^{-1} & \text{ for $ij\in BB$};\\
1 & \text{ for $ij\in AB$ or $BA$}.
\end{array}
\right.
\end{eqnarray}
Hence, a binary system of $N$ particles with non-reciprocal interactions of the form of Eq.~(\ref{force_const}) is described
by a {\it pseudo-Hamiltonian} with the masses (\ref{mass_ren}) and interactions (\ref{phi_ren}). In particular, this implies
the pseudo-momentum and energy conservation,
\begin{eqnarray*}
\sum\limits_{i}^N\tilde m_i{\bf v}_i &=& {\rm const}, \\
\sum\limits_{i}^N\frac12\tilde m_iv_i^2+\sum\limits_{i<j}^N\tilde\varphi(r_{ij}) &=& {\rm const},
\end{eqnarray*}
and allows us to employ the methods of equilibriums statistical mechanics to describe such systems. For instance, from
equipartition, $\frac12\tilde m_A\langle v_A^2\rangle=\frac12\tilde m_B\langle v_B^2\rangle\equiv\frac12Dk_{\rm B}\tilde T$
(where $\tilde T$ is the pseudo-temperature and $D$ is the dimensionality), it immediately follows that in detailed balance
$T_A=(1+\Delta)\tilde T$ and $T_B=(1-\Delta)\tilde T$, i.e.,
\begin{equation}\label{equilibrium}
\frac{T_A}{T_B}=\frac{1+\Delta}{1-\Delta}.
\end{equation}
We conclude that a mixture of particles with non-reciprocal interactions can be in a remarkable state of equilibrium, where
the species have different temperatures $T_A$ and $T_B$. Such equilibrium is only possible for $\Delta<1$, otherwise the
forces ${\bf F}_{AB}$ and ${\bf F}_{BA}$ are pointed in the {\it same} direction [see Eq.~(\ref{force_const})] and the
system cannot be stable.

Now we shall study a general case, when the interaction between the species $A$ and $B$ is determined by the force whose
reciprocal, ${\bf F}_{\rm r}(r)$, and non-reciprocal, ${\bf F}_{\rm n}(r)$, parts are arbitrary functions of the
interparticle distance $r$.
Both forces can be presented as ${\bf F}_{\rm r,n}(r)=({\bf r}/r)F_{\rm r,n}(r)$, where $F_{\rm r,n}=-d\varphi_{\rm
r,n}/dr$. It is instructive to write the equations of motion for a pair of interacting particles $A$ and $B$ in terms of the
relative coordinate ${\bf r}= {\bf r}_{A}-{\bf r}_{B}$ and the center-of-mass coordinate ${\bf R}= (m_A{\bf r}_{A}+m_B{\bf
r}_{B})/M$,
\begin{eqnarray}
M\ddot{\bf R} &=& 2{\bf F}_{\rm n}(r),\label{motion_C}\\
\mu\ddot{\bf r} &=& {\bf F}_{\rm r}(r)+\frac{m_B-m_A}{m_A+m_B}{\bf F}_{\rm n}(r),\label{motion_R}
\end{eqnarray}
where $\mu= m_Am_B/(m_A+m_B)$ and $M= m_A+m_B$ are the reduced and total masses, respectively. We define the relative
velocity, ${\bf v}= \dot{\bf r}$, the center-of-mass velocity, ${\bf V}= \dot{\bf R}$, and their values after a collision,
${\bf v}'={\bf v}+\delta{\bf v}$ and ${\bf V}'={\bf V}+\delta{\bf V}$. From Eq.~(\ref{motion_R}) we conclude that the
relative motion is conservative, i.e., the absolute value of the relative velocity remains unchanged after a collision,
$|{\bf v}+\delta{\bf v}|=|{\bf v}|$. Equation~(\ref{motion_C}) governs the variation of the center-of-mass velocity,
$\delta{\bf V}$, which is determined by the relative motion via ${\bf F}_{\rm n}(r)$. By employing the relation ${\bf
v}_{A,B}={\bf V}\pm(\mu/m_{A,B}){\bf v}$, we obtain the variation of the kinetic energy $E_{A,B}$ after a collision:
\begin{eqnarray}
\delta E_{A,B}=m_{A,B}\left[{\bf V}\cdot\delta{\bf V}+\frac12(\delta{\bf V})^2\right]\hspace{2.cm}\nonumber\\
\pm\mu\left({\bf V}\cdot\delta{\bf v}+{\bf v}\cdot\delta{\bf V}+\delta{\bf V}\cdot\delta{\bf v}\right).\label{delta_E}
\end{eqnarray}
Let us introduce the angle $\theta$ between ${\bf V}$ and ${\bf v}$, and the scattering angle $\chi$ between ${\bf v}'$ and
${\bf v}$. Since the relative motion is conservative, from Eq.~(\ref{motion_C}) we conclude that $\delta{\bf V}$ is parallel
to $\delta{\bf v}$. Hence, $\delta{\bf V}\cdot\delta{\bf v}=\delta V\delta v$, and for two-dimensional (2D) systems we have
${\bf V}\cdot\delta{\bf V}=V\delta V\sin(\theta-\frac12\chi)$, ${\bf V}\cdot\delta{\bf v}=V\delta
v\sin(\theta-\frac12\chi)$, and ${\bf v}\cdot\delta{\bf V}=-v\delta V\sin\frac12\chi$
\cite{note1,LandauMechanics,LandauKinetics}.

In order to calculate the magnitudes of the velocity variations and the scattering angle, we consider the {\it small-angle}
scattering, $\chi\ll1$ \cite{LandauMechanics}: Such approximation significantly simplifies the general analysis and is valid
for sufficiently high temperatures (provided the pair interaction is not of the hard-sphere-like type). Using
Eqs.~(\ref{motion_C}) and (\ref{motion_R}), for a given impact parameter $\rho$ we get $\delta V(\rho)=(4/M v)f_{\rm
n}(\rho)$ and $\chi(\rho)=\delta v/v=(2/\mu v^2)[f_{\rm r}(\rho)+ \frac{m_B-m_A}{m_A+m_B}f_{\rm n}(\rho)]$, expressed via
the scattering functions ($\alpha=$r,n):
\begin{equation*}
f_{\alpha}(\rho)=\rho\int_{\rho}^{\infty}dr\frac{F_{\alpha}(r)}{\sqrt{r^2-\rho^2}}.
\end{equation*}

The equations describing evolution of the mean kinetic energy of the species $A$ and $B$ can be obtained by multiplying
$\delta E_{A,B}$ with the collision frequency between the species and averaging it over the velocity distributions
\cite{LandauKinetics}. The collision cross section is represented by the integral over the impact parameter
\cite{LandauMechanics}, $\int d\rho$ for 2D systems or $\int d\rho\:2\pi\rho$ for 3D systems. To obtain a closed-form
solution, we shall assume that the elastic momentum/energy exchange in collisions provides efficient Maxwellization of the
distribution functions (which can be verified by molecular dynamics simulations, see below). Then, one can perform the
velocity averaging over the Maxwellian distributions with the temperatures $T_{A,B}$ (note that after the integration over
$\theta$ all terms in Eq.~(\ref{delta_E}) yield contributions $\sim\chi^2$). After some algebra we derive the following
equations for 2D systems:
\begin{widetext}
\begin{equation}\label{kinetics}
\dot T_{A,B}=\pm\frac{1\pm\Delta_{\rm eff}}{1+\epsilon}\frac{\sqrt{2\pi}n_{B,A} I_{\rm rr}}{m_Am_B\left(\frac{T_A}{m_A}
+\frac{T_B}{m_B}\right)^{3/2}}\left[(1+\Delta_{\rm eff})T_B-(1-\Delta_{\rm eff})T_A+\frac{\epsilon}{1\pm\Delta_{\rm eff}}
(T_B-T_A)\right],
\end{equation}
\end{widetext}
where $n_{\alpha}$ is the areal number density (for simplicity, below we assume $n_A=n_B=n$). The equations depend on the
{\it effective non-reciprocity} $\Delta_{\rm eff}$ and the {\it interaction disparity} $\epsilon$,
\begin{equation}\label{parameters}
    \begin{array}{l}
    \Delta_{\rm eff} = I_{\rm nn}/I_{\rm rn},\\[.2cm]
    \epsilon = I_{\rm rr}I_{\rm nn}/I_{\rm rn}^2-1,
    \end{array}
\end{equation}
expressed via the integrals $I_{\alpha\beta}=\int_0^{\infty}d\rho\: f_{\alpha}f_{\beta}$ (naturally, it is assumed that the
integrals converge). We point out that $\Delta_{\rm eff}$ and $\epsilon$ are numbers uniquely defined for given functions
$\varphi_{\rm r,n}(r)$; from the Cauchy inequality it follows that $\epsilon\geq0$.

Note that for 3D systems the r.h.s. of Eq.~(\ref{kinetics}) should be multiplied by the additional factor 8/3, and the
integrals become $I_{\alpha\beta}=\int_0^{\infty}d\rho\: \rho f_{\alpha}f_{\beta}$. For a reciprocal Coulomb interaction,
$I_{\rm rr}$ is proportional to
the so-called Coulomb logarithm (see e.g., \cite{LandauKinetics,Spitzer}) and $\Delta_{\rm eff}=0$. In this case
Eq.~(\ref{kinetics}) is reduced to the classical equation for the temperature relaxation in a plasma \cite{LandauKinetics}.

\begin{figure}
\includegraphics[width=.99\columnwidth,clip=]{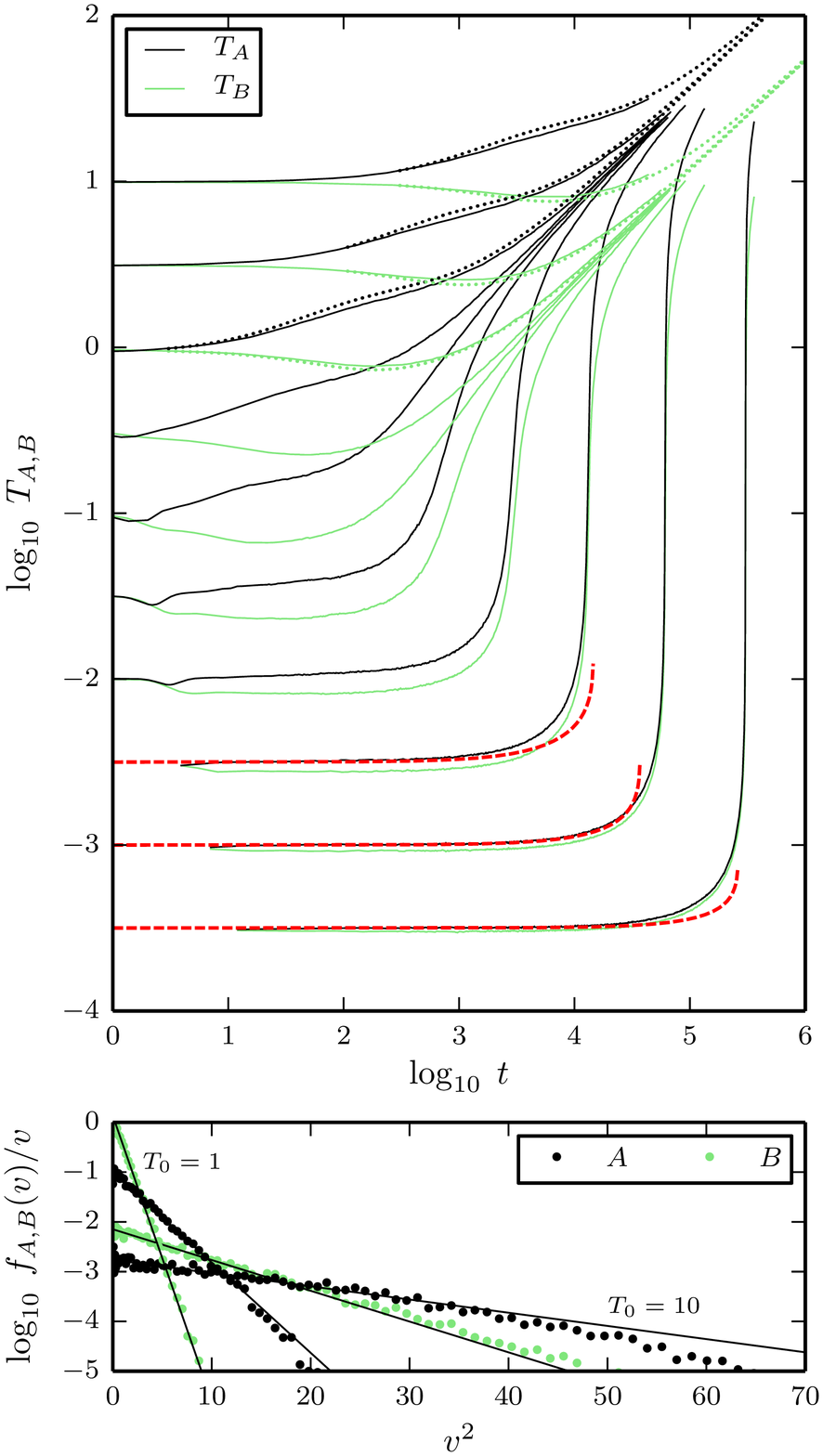}
\caption{Growth of the mean kinetic energy in a 2D binary system of particles interacting via the non-reciprocal Hertzian
forces (see text for details). The upper panel displays the time dependence of the temperatures $T_A$ and $T_B$. The solid
lines show the development obtained from the simulations for the areal fraction $\phi=\pi r_0^2n=0.3$ and different initial
temperatures $T_0$. All curves approach the
universal asymptotes $\propto t^{2/3}$ described by Eqs.~(\ref{asymp1}) and (\ref{asymp2}). The doted lines represent the
solution of Eq.~(\ref{kinetics}) for $T_0\gtrsim1$. The early development at $T_0\ll1$ is fitted by the explosive solution,
Eq.~(\ref{explosion}), shown by the dashed lines. The temperatures are normalized by $\varphi_0$, time is in units of
$\sqrt{mr_0^2/\varphi_0}$. The lower panel illustrates the velocity distributions $f_{A,B}(v)$ at $t\simeq700$ for $T_0=1$
and 10.
}\label{Fig.1}
\end{figure}

For the ``constant'' non-reciprocity, $F_{\rm n}(r)/F_{\rm r}(r)=\Delta$, we get $\Delta_{\rm eff}=\Delta$ and $\epsilon=0$.
In this case Eq.~(\ref{kinetics}) yields the equilibrium $\dot T_{A,B}=0$ for the temperature ratio given by
Eq.~(\ref{equilibrium}). Otherwise we have $\epsilon>0$ and the temperatures grow with time, approaching the asymptotic
solution,
\begin{equation}\label{asymp1}
t\to\infty:\quad T_A(t)=\tau T_B(t)=ct^{2/3},
\end{equation}
where $c\propto(nI_{\rm rr})^{2/3}$ and
\begin{equation}\label{asymp2}
\tau=\sqrt{\frac{(1+\Delta_{\rm eff})^2+\epsilon}{(1-\Delta_{\rm eff})^2+\epsilon}}.
\end{equation}
Thus, the asymptotic temperature ratio is a constant which tends to the equilibrium value [Eq.~(\ref{equilibrium})] for
$\epsilon\to0$.

\begin{figure}
\includegraphics[width=\columnwidth,clip=]{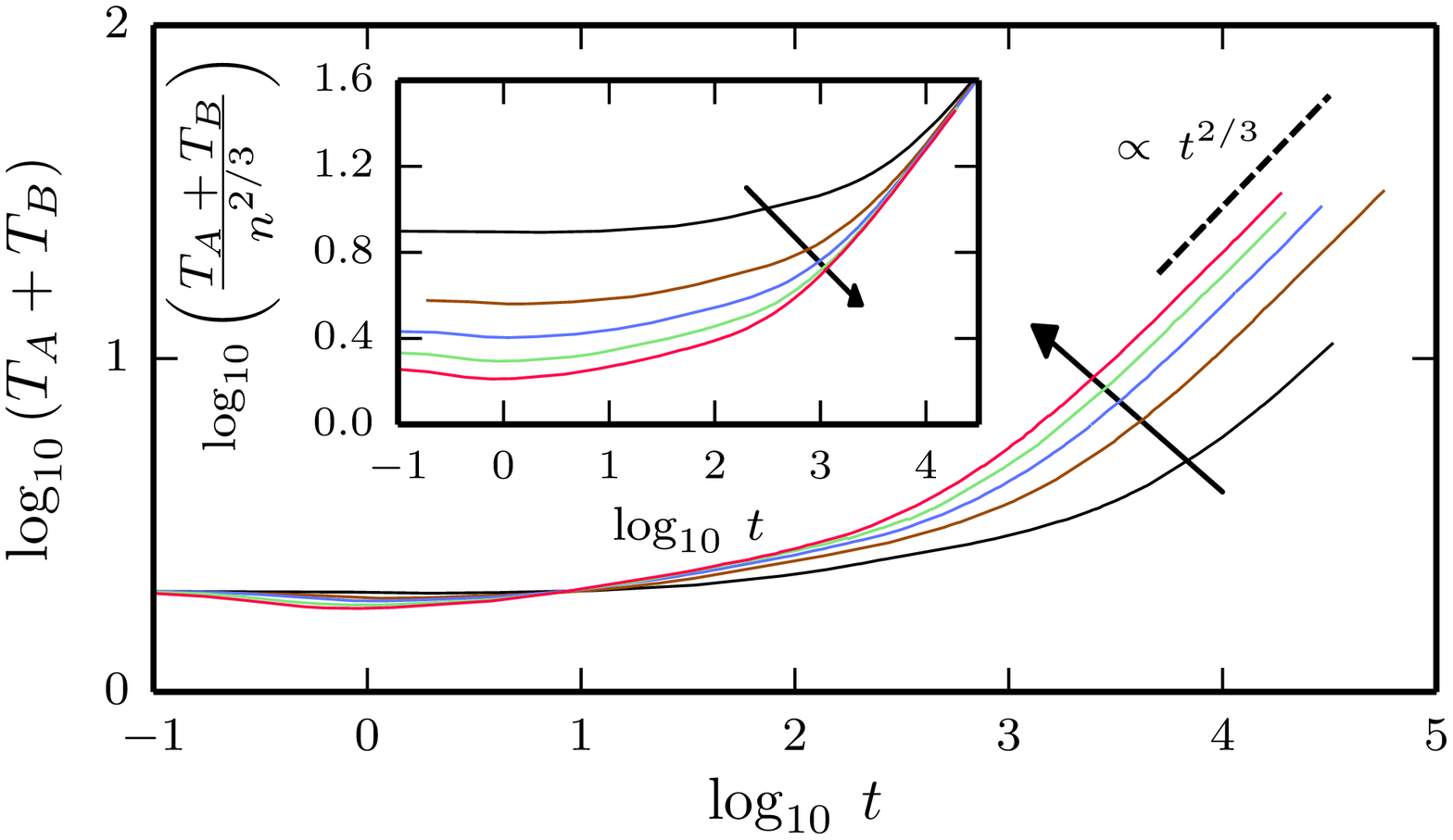}
\caption{Time dependence of the total kinetic energy, $T_A+T_B$, obtained from the simulations. The development for $T_0=1$
and different $\phi$ (0.1, 0.3, 0.5, 0.7, 0.9, increasing along the arrow) is shown. The inset demonstrates the
$\propto n^{2/3}$ density scaling of the asymptotic temperature growth. The temperature and time units are the same as
in Fig.~\ref{Fig.1}.}
\label{Fig.2}
\end{figure}

To verify the analytical results, we carried out a molecular dynamics simulation of a 2D binary, equimolar mixture of soft
spheres. We implemented the velocity Verlet algorithm \cite{Swope82} with an adaptive time step. The simulation box with
periodic boundary conditions contained $2\times20,000$ particles with equal masses. We used a model Hertzian potential
\cite{note2,Pamies09,Berthier10} whose reciprocal and non-reciprocal parts are given by $\varphi_{\rm
r}(r)=\frac12\varphi_0(\max\{0, 1-r/r_0\})^2$ and $\varphi_{\rm n}(r)=\frac13\varphi_0(\max\{0, 1-r/r_0\})^3$, respectively,
where $\varphi_0$ is the interaction energy scale and $r_0$ is the interaction range. At $t=0$ the particles were arranged
into two interpenetrating square lattices with the initial temperature $T_A = T_B = T_0$ (therefore, at early simulation
time a certain fraction of $T_0$ was converted into the interaction energy).

The numerical results are illustrated in Fig.~\ref{Fig.1}, where we plot the dependencies $T_{A,B}(t)$ for different $T_0$.
For the Hertzian interactions, from Eq.~(\ref{parameters}) we obtain $\Delta_{\rm eff}=0.57$ and $\epsilon=0.082$, and
Eq.~(\ref{asymp2}) yields the asymptotic temperature ratio $\tau=3.1$. One can see that for all $T_0$ the numerical curves
approach the expected universal asymptotes described by Eqs.~(\ref{asymp1}) and (\ref{asymp2}). Note that the early
development at sufficiently low temperatures exhibits a remarkably sharp dependence on $T_0$ -- we observe the formation of
a plateau which broadens dramatically with decreasing $T_0$. On the other hand, for $T_0\gtrsim1$ the numerical results are
very well reproduced by the solution of Eq.~(\ref{kinetics}), as expected. A small ($<10\%$) deviation observed in this case
is due to the fact that weak collisions are no longer providing efficient Maxwellization of the velocity distribution for
the ``hotter'' species $A$ (see the lower panel of Fig.~\ref{Fig.1}).

In Fig.~\ref{Fig.2} we show how the temperature evolution depends on the density $n$. Here, the total kinetic energy
$T_A(t)+T_B(t)$ calculated for different values of the areal fraction $\phi=\pi r_0^2n$ is plotted. In contrast to the sharp
dependence on $T_0$ seen in Fig.~\ref{Fig.1}, the increase of $n$ is accompanied by an approximately proportional shortening
of the plateau \cite{note3}. The inset demonstrates the predicted $\propto n^{2/3}$ scaling for the asymptotic temperature
growth.

In order to explain the observed behavior at low temperatures, we point out that the approximation of small-angle scattering
is not applicable in this regime and, hence, Eq.~(\ref{kinetics}) is no longer valid. Strong correlations make the analysis
rather complicated in this case, but one can implement a simple phenomenological model to understand the essential features.
We postulate that at sufficiently low temperatures the energy growth caused by non-reciprocal interactions can be balanced
by nonlinearity, forming a ``dynamical potential well'' where the system can reside for a long time. Qualitatively, one can
then expect the development around the initial temperature to be governed by the activation processes, and introduce the
effective ``Arrhenius rate'' characterizing these processes. Assuming the dimensionless temperature $T$ (normalized by the
effective depth of the well) to be small, we employ the following model equation:
\begin{equation}\label{initial}
\dot T=C\exp(T^{-\gamma}),
\end{equation}
where $C$ is a constant (possible power-law factors can be neglected for $T\ll1$) and $\gamma$ is an exponent determined by
the particular form the potential well. Substituting $T^{-\gamma}\simeq T_0^{-\gamma}-\gamma T_0^{-\gamma-1}(T-T_0)$ in
Eq.~(\ref{initial}) yields the explosive solution,
\begin{equation}\label{explosion}
T(t)=T_0-\frac{T_0^{\gamma+1}}{\gamma}\ln\left[1-\frac{C\gamma}{T_0^{\gamma+1}}\exp(-T_0^{-\gamma})t\right],
\end{equation}
with the explosion time $t_{\rm ex}=(T_0^{\gamma+1}/C\gamma)\exp(T_0^{-\gamma})$. In Fig.~\ref{Fig.1} we show that the
explosive solution provides quite a reasonable fit to the numerical results at low temperatures for $C=4\times10^{-5}$ and
$\gamma=0.305$.

Let us briefly discuss the effect of dissipation due to friction against the surrounding medium. To take this into account,
one has to add the dissipation term $-2\nu_{A,B}(T_{A,B}-T_{\rm b})$ to the r.h.s. of Eq.~(\ref{kinetics}), where
$\nu_{\alpha}$ is the respective damping rate in the friction force $-m_{\alpha}\nu_{\alpha}{\bf v}_{\alpha}$ and $T_{\rm
b}$ is the background temperature (determined by the interaction of individual particles with the medium)
\cite{Morfill09,vanKampen}. In this case the temperatures $T_{A,B}$ always reach a steady state, since the growth term in
Eq.~(\ref{kinetics}) decreases with temperature. The resulting steady-state temperature ratio, $\tau_{\nu}$, can be easily
derived from Eq.~(\ref{kinetics}), assuming that the steady-state temperatures are much larger than $T_{\rm b}$. For similar
particles, this requires the condition $\nu\ll nI_{\rm rr}/\sqrt{mT_{\rm b}^3}$ to be satisfied \cite{note4}. Then we obtain
the following equation for $\tau_{\nu}$:
\begin{equation*}
\tilde\nu[(1-\Delta_{\rm eff})^2+\epsilon]\tau_{\nu}^2-(\tilde\nu-1)(1-\Delta_{\rm eff}^2+\epsilon)\tau_{\nu}
=(1+\Delta_{\rm eff})^2+\epsilon,
\end{equation*}
where $\tilde\nu=\nu_A/\nu_B$. For $\tilde\nu=1$ we get $\tau_\nu=\tau$, i.e., the steady-state temperature ratio is not
affected by friction. Generally, $\tau_{\nu}$ exhibits a very weak dependence on $\tilde\nu$: e.g., for the Hertzian
interactions the deviation between $\tau_{\nu}$ and $\tau$ is within $\simeq1\%$ in the range $0.8\leq\tilde\nu\leq1.3$
(expected for experiments with binary complex plasmas \cite{Morfill09,Comm}).

Note that at low temperatures the system can be dynamically ``arrested'' due to friction and never reach the asymptotic
stage described by Eqs.~(\ref{asymp1}) and (\ref{asymp2}). A simple analysis of Eq.~(\ref{initial}) with the dissipation
term shows that the arrest occurs when $\nu t_{\rm ex}\gtrsim1$.

In conclusion, the presented results provide a basic classification of many-body systems with non-reciprocal interactions.
We investigated different non-reciprocity classes in 2D and 3D systems which are relevant to a plethora of real situations:
For instance, the shadow interactions \cite{Tsytovich97,Chaudhuri11} in binary complex plasmas have a constant
non-reciprocity and can dominate the kinetics of 3D systems, while the wake-mediated interactions \cite{Morfill09,Couedel10}
governing the action-reaction symmetry breaking in bilayer complex plasmas are generally characterized by a variable
non-reciprocity. We expect that our predictions can be verified in complex plasma experiments, e.g., by measuring the
kinetic temperatures in 2D binary mixtures or in 3D clouds under microgravity conditions. Furthermore, the analysis of
dynamical correlations in the strong-damping regime should help to understand the effect of non-reciprocal effective
interactions operating in colloidal suspensions.

The authors acknowledge support from the European Research Council under the European Union's Seventh Framework Programme,
Grant Agreement No. 267499.


\begin{thebibliography}{99}
\bibitem{note0} For the $AB$ and $BA$ pairs $\varphi(r)$ is the same, to distinguish the effect of non-reciprocity.
\bibitem{note1} For 3D systems the corresponding expressions are easily derived using the cosine rule of spherical
    trigonometry.
\bibitem{note2} We chose the Hertzian forces for the illustration, in order to ensure precise numerical calculations
    at low and high temperatures.
\bibitem{note3} A small dip in the early development (Fig.~\ref{Fig.2}) is due to partial conversion of the initial kinetic
    energy into the interaction energy.
\bibitem{note4} This strong inequality always holds for typical experiments with 2D complex plasmas, where the interparticle
    interaction can be approximated by the Yukawa potential, $\varphi_{\rm r}(r)=(Q^2/r)e^{-r/\lambda}$, so that
    $I_{\rm rr}\sim Q^4/\lambda$.
\end{thebibliography}
\end{document}